\documentclass[prb,aps,twocolumn,showpacs,ams]{revtex4}
\usepackage{graphicx} 
\usepackage{tabularx}
\usepackage{dcolumn} 
\usepackage{color}
\usepackage{bm}
\usepackage{amsmath}
\usepackage{amssymb}

\begin{document}



\title{
Interplay between effective mass anisotropy and Pauli paramagnetic effects in a multiband superconductor
 ---Application to Sr$_2$RuO$_4$---
}



\author{Noriyuki Nakai$^1$,  and Kazushige Machida$^{1,2}$} 
\affiliation{
$^1$Department of Physics, Okayama University, Okayama 700-8530, JAPAN, 
$^2$Department of Physics, Ritsumeikan University, Kusatsu 525-8577, JAPAN
}




\date{\today}

\begin{abstract}
We investigate the mixed state properties in a type II multiband superconductor
with uniaxial anisotropy under the Pauli paramagnetic effects.
Eilenberger theory extended to a multiband superconductor is utilized to describe the detailed vortex lattice properties,
such as the flux line form factors, the vortex lattice anisotropy and magnetic torques.
We apply this theory to Sr$_2$RuO$_4$ to analyze those physical quantities obtained experimentally,
focusing on the interplay between the strong two-dimensional anisotropy and the 
Pauli paramagnetic effects.
This study allows us to understand the origin of the disparity between the vortex lattice anisotropy ($\sim$60) 
and the $H_{\rm c2}$ anisotropy ($\sim$20). Among the three bands; $\gamma$  with the effective mass anisotropy
$\sim$180,  $\alpha$ with $\sim$120, and $\beta$ with $\sim$60, the last one is found to be the major band, responsible
for various magnetic responses while the minor $\gamma$ band plays an important role in the vortex formation.
Namely, in a field orientation slightly tilted away from the two dimensional basal plane those two bands cooperatively
form the optimal vortex anisotropy which exceeds that given by the effective mass formula with infinite anisotropy.
This is observed by small angle neutron scattering experiments on Sr$_2$RuO$_4$. 
The pairing symmetry of Sr$_2$RuO$_4$ realized is either spin singlet or spin triplet with the d-vector strongly locked in the basal plane.
The gap structure is that the major $\beta$ band has a full gap and the minor $\gamma$ band has a $d_{x^2-y^2}$ like gap.
\end{abstract}



\maketitle

\section{Introduction}
It is now widely recognized that multiband superconductors are omnipresent \cite{perali}.
This recognition may be triggered by MgB$_2$ \cite{akimitsu} where there exist distinctive two bands;
the 3D $\pi$-band and 2D like $\sigma$-band\cite{mazin}.
They play different role in forming superconductivity, in particular in magnetic
properties under an applied field, such as symmetry of vortex lattices\cite{hirano} or the form factors
probed by small angle neutron scattering (SANS) experiments\cite{mortenmgb2}.
To understand its detailed magnetic response, a two band model is indispensable.
In fact different dimensionality of the band structures between  the $\pi$-band and $\sigma$-band gives
rise to rotation of the triangular vortex lattice under varying field\cite{hirano}.
The form factors of SANS experiments clearly demonstrate a gradual change of the two components
of  the $\pi$-band and  $\sigma$-band as field varies\cite{mortenmgb2}.
It is also true for other materials among unconventional and conventional superconductors where
multiband description is essential, such as heavy fermion superconductors~\cite{kittakace,shimizu} and iron pnictides~\cite{kittaka122,kuhn}.

We have been seeing that the Pauli paramagnetic effect (PPE) is important when combined with this multiband effect
in certain superconductors, which give rise to a variety of unexpected phenomena.
Typical examples are the oldest heavy fermion superconductors CeCu$_2$Si$_2$\cite{kittakace} and UBe$_{13}$~\cite{shimizu},
and also KFe$_2$As$_2$~\cite{kittaka122,kuhn}, which
necessitate the multiband description in fully understanding of their vortex properties.
Those include the hidden first order transition phenomenon~\cite{tsutsumi} and the disparity~\cite{kuhn} between the vortex lattice anisotropy
and Fermi velocity anisotropy as discussed later.

\begin{figure}
\begin{center}
\includegraphics[width=9cm]{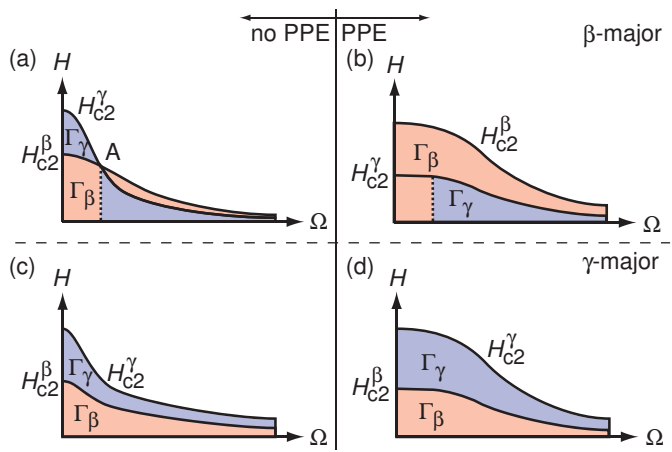}
\end{center}
\caption{\label{fig1}(Color online)
Schematic phase diagrams in the $H$ versus $\Omega$ plane. The $\beta$ band major scenario without
PPE (a) and with PPE (b). The $\gamma$ band major scenario without
PPE (c) and with PPE (d). The effective mass anisotropy is assumed to be $\Gamma_{\gamma}>\Gamma_{\beta}$.
}
\end{figure}

Here we study the interplay between PPE and multiband effects in a uniaxial anisotropic
superconductor in which each band has a different uniaxial anisotropy:
For the case of two bands which we consider in this paper we can envisage two possible situations as schematically 
illustrated in Fig.~\ref{fig1}.
Let us consider the first case shown in Figs.~\ref{fig1} (a) and (b). In the absence of PPE the two orbital limited upper critical fields
$H^{\rm orb}_{\rm c2, \gamma}$ and
$H^{\rm orb}_{\rm c2, \beta}$ cross at A in $H$ versus $\Omega$ plane ($\Omega$ is the angle from the $ab$ plane),
each of which is characterized by the effective mass anisotropy $\Gamma_i$ for two bands $i=\gamma$ and $\beta$,
assuming $\Gamma_{\gamma}>\Gamma_{\beta}$.
As indicated in Fig.~\ref{fig1} (a) the four divided regions are characterized by each $\Gamma_i$.
In particular, along $H_{\rm c2}(\Omega)$ the characteristic anisotropy of the total system 
is switched at the intersecting point A from
$\Gamma_{\gamma}$ to $\Gamma_{\beta}$ as $\Omega$ increases.
When traversing at a higher $H$,  the anisotropy $\Gamma_{\gamma}$ for the $\gamma$ band
is sensed only.

Now let us switch on PPE, then both orbital limited $H^{\rm orb}_{\rm c2}(\Omega)$ are suppressed towards lower fields,
especially $\Gamma_{\gamma}$ if we assume that the superconducting gaps for the two bands
such that $\Delta_{\beta}>\Delta_{\gamma}$. This is because the Pauli limited fields $H_p^{\beta}>H_p^{\gamma}$.
The resulting phase diagram is shown in Fig.~\ref{fig1} (b).
As seen from it the crossing point A is removed and three regions are now occupied by  $\Gamma_{\beta}$
while the $\Gamma_{\gamma}$ region is hidden deep inside at lower fields and finite $\Omega$'s.
In particular, along $H_{\rm c2}(\Omega)$ the $\Gamma_{\beta}$ region persists all the way from $\Omega=0^{\circ}$
 to $\Omega=90^{\circ}$. Thus the higher field scan is sensing only the $\Gamma_{\beta}$ anisotropy 
 while a lower field scan is sensing $\Gamma_{\beta} \rightarrow \Gamma_{\gamma} \rightarrow \Gamma_{\beta}$
 as $\Omega$ increases. This non-trivial anisotropy evolution is caused 
 by the interplay between the effective mass anisotropy 
$\Gamma$ and PPE.

Another possible phase diagrams are depicted in Figs.~\ref{fig1}(c) and (d) where $\Delta_{\beta}<\Delta_{\gamma}$ is assumed,
keeping $\Gamma_{\gamma}>\Gamma_{\beta}$.
The $\gamma$ ($\beta$) band with $\Delta_{\gamma}$ ($\Delta_{\beta}$) is now major (minor).
In the absence of PPE shown in Fig.~\ref{fig1}(c) the two orbital limited $H^{\rm orb}_{\rm c2}(\Omega)$ 
curves are not crossed, thus the higher (lower)
field region is occupied by $\Gamma_{\gamma}$($ \Gamma_{\beta}$).
In the presence of PPE those curves are both suppressed downwards.
Thus the two regions ($\Gamma_{\gamma}$ and $ \Gamma_{\beta}$ ) are simply shifted downwards, 
keeping its phase diagram topologically
unchanged. Note that $H_p^{\beta}<H_p^{\gamma}$. 
Therefore, a higher (lower) field scan is exclusively sensing the $\Gamma_{\gamma}$ ($ \Gamma_{\beta}$) anisotropy
of the total system as $\Omega$ increases.

The purpose of this paper is to investigate the physics of Sr$_2$RuO$_4$ through the studies of
the mixed state properties in this multiband  superconductor
where the interplay between the effective mass (or Fermi velocity) anisotropy  and 
PPE is important in understanding of the vortex lattice state in $H$ vs $\Omega$ plane.
This superconductor is known to have the 
strong two dimensional uniaxial anisotropies for the three bands~\cite{MackenzieMaeno}, the $\gamma$ band ($\Gamma_{\gamma}\cong$180)
the $\alpha$ band ($\Gamma_{\alpha}\cong$120) and the $\beta$ band ($\Gamma_{\beta}\cong$60).
Since the density of states (DOS) $N_{{\rm F}i}$ at the Fermi level ($N_{{\rm F}\gamma}$=0.53, $N_{{\rm F}\alpha}$=0.10 and $N_{{\rm F}\beta}$=0.37
of the total DOS), the $\alpha$ band is neglected in this paper for simplicity.
In the following we consider a two band model: the $\gamma$ band with $N_{{\rm F}\gamma}$=0.53
and the $\beta$ band with $N_{{\rm F}\beta}$=0.47.

In Sr$_2$RuO$_4$ there are several outstanding unsolved issues:

\noindent
(1) Which bands is the major band for superconductivity, either the $\gamma$ band or the $\beta$ band ?
$\Delta_{\beta}>\Delta_{\gamma}$ or $\Delta_{\beta}<\Delta_{\gamma}$?

\noindent
(2) Can the PPE resolve the observed disparity between the vortex lattice anisotropy~\cite{Rastovski} 60
and the $H_{\rm c2}$ anisotropy~\cite{MackenzieMaeno} 20 ?

\noindent
The first issue (1) has been extensively debated, in particular in connection with 
the pairing mechanism of this material to stabilize the chiral p-wave state~\cite{zhito,nomura,kivelson,simon}.
We approach this issue from the different view point by analyzing the mixed state vortex states.
The second issue (2) has been investigated by us~\cite{Machida,amano2} based on the single band picture.
In this paper we revisit it based on a more realistic two band model, which is able to allow us to 
study the issue (1).

We base our computations on the quasiclassical theory~\cite{eilenberger,larkin1,usadel}.
The original single band theory is extended to various multiband cases, including
MgB$_2$~\cite{koshelev,nakai2band}, iron pnictides~\cite{vorontsov,schmalian,efetov}.
The applicability of the quasiclassical theory~\cite{eilenberger,larkin1,usadel} is given in general
by the condition $k_{\rm F}\xi\gg 1$  with $k_{\rm F}$ the Fermi wave number and 
$\xi$ the coherence length under the assumption that the normal state properties are described by
a Fermi liquid theory.
For Sr$_2$RuO$_4$ the three bands $\alpha$, $\beta$ and $\gamma$
are $k_{\rm F}$=0.304, 0.622 and 0.753 ($\AA^{-1}$), respectively.
The in-plane and $c$ axis coherence lengths $\xi$ are 660 and 33 ($\AA$), respectively~\cite{MackenzieMaeno}.
Thus for any combinations $k_{\rm F}\xi\gg 1$ is well satisfied.
We also notice that the lattice constants $a$=3.862$\AA$ and $c$=12.722$\AA$
are short enough compared with those coherence lengths, thus Sr$_2$RuO$_4$
is a three dimensional normal metal.
Moreover, most physical quantities, including various transport coefficients and thermodynamic
properties can be consistently and coherently  described by a Fermi liquid theory
as explained in details by Mackenzie and Maeno~\cite{MackenzieMaeno}.
Therefore we can quite safely apply the quasiclassical theory to Sr$_2$RuO$_4$.

By self-consistently solving microscopic quasiclassical Eilenberger equation with two bands,
we calculate a variety of the physical quantities relevant to available experiments,
such as the form factors probed by SANS experiments~\cite{Rastovski,morten},
magnetic torques~\cite{kittaka}, and the vortex lattice anisotropy $\Gamma_{\rm VL}$ 
which differs generally from the effective mass anisotropy $\Gamma_i$ 
mentioned above.

The arrangement of this paper is as follows:
After introducing Eilenberger theory extended to multibands~\cite{nakai2band}  with PPE in Sec. 2
which was done in our previous paper~\cite{tsutsumi}, we construct a model 
system for considering Sr$_2$RuO$_4$ by fixing the several model parameters in Sec. 3.
Here we compare the two scenarios, one is the $\beta$ major and the other $\gamma$
in equal footing, finding that the former is better than the latter relative to the existing	
experiments. 
Then we come to the main theme of the present paper in Sec.4; 
computations and analyses of the form factors and vortex lattice anisotropies
as a function of the angle $\Omega$ compared with the data of SANS experiments~\cite{Rastovski,kuhn}.
In Sec. 5 we examine the magnetic torques which are also measured recently by Kittaka et al\cite{kittaka}.
The final section 6 is devoted to discussions and conclusion.
This paper is an extension of our previous work based on a single band model\cite{amano2,amano1,ishihara}
and also closely related to our two band model calculations\cite{tsutsumi,nakai2band}.

\section{Quasiclassical theory including PPE for two bands}
\label{sec:formulation}

We start with the free energy $F$ in the quasiclassical theory~\cite{eilenberger,larkin1,usadel} 
extended to the two band case~\cite{koshelev,nakai2band,vorontsov,schmalian,efetov}, which is given by 
\begin{equation}
F=\int d\bm{r}\{
\frac{|B(\bm{r})|^2}{8\pi}-\frac{\chi_n}{2}|B(\bm{r})|^2+\Sigma_{i,j}\Delta^{\ast}_j(\bm{r})({\hat V}^{-1})_{i,j}
\Delta_i(\bm{r})
\nonumber
\end{equation}
\begin{equation}
-\pi k_{\rm B}TN_{{\rm F}0}\Sigma_{|\omega_n|<\omega_c}\Sigma_{j}\frac{N_{{\rm F}j}}{N_{{\rm F}0}}\langle I(\omega_n,\bm{k}_j,\bm{r})\rangle_{\bm{k}_j}
\}
\label{eq:freeenergy}
\end{equation}
\noindent
with $\chi_n=2\mu^2_{\rm B}N_{{\rm F}0}$, $N_{{\rm F}0}=\Sigma_jN_{{\rm F}j}$ and
\begin{equation}
I(\omega_n,\bm{k}_j,\bm{r})=\Delta(\bm{k}_j,\bm{r})f^{\dagger}(\omega_n,\bm{k}_j,\bm{r})
+\Delta^{\ast}(\bm{k}_j,\bm{r})f(\omega_n,\bm{k}_j,\bm{r})
\nonumber
\end{equation}
\noindent
\begin{equation}
+(g_j-sgn(\omega_n))\{
\frac{1}{f_j}(\omega_n+i\mu_{\rm B}B+\frac{\hbar}{2}{\bm v}_{{\rm F}j}\cdot ({\vec\nabla}-i\frac{2\pi}{\phi_0}\bm A))f_j
\nonumber
\end{equation}
\begin{equation}
+\frac{1}{f_j^{\dagger}}(\omega_n+i\mu_{\rm B}B-\frac{\hbar}{2}{\bm v}_{{\rm F}j}\cdot ({\vec\nabla}+i\frac{2\pi}{\phi_0}\bm A))f_j^{\dagger}
\}
\label{eq:freeenergy2}
\end{equation}
\noindent
The flux quantum $\phi_0=\frac{hc}{2|e|}$. ${\bm v}_{{\rm F}j}$ is the Fermi velocity at ${\bm k}_{j}$ of the  band j.
The Fermi surface average $\langle \cdots \rangle_{{\bf k}_j}$ is normalized within each band as
$\langle 1\rangle_{{\bf k}_j}=1$.
Here we introduced the interaction matrix $V_{ij}$ with 2$\times$2 for two bands
where $V_{jj}$ is the pairing interaction on the $j$ band and $V_{ij}=V_{ji}$ for $i\neq j$ is the Cooper pair transfer between
the $i$ and $j$ bands.
$g_j=g(\omega_n,\bm{k}_j,\bm{r})$, $f_j=f(\omega_n,\bm{k}_j,\bm{r})$, and $f^{\dagger}_j=f^{\dagger}(\omega_n,\bm{k}_j,\bm{r})$
are the quasiclassical Green's functions for the $j$ band.
$\Delta_j(\bm{k}_j,\bm{r})=\Delta_j(\bm{r})\phi_j(\bm{k}_j)$ is the pair potential and $\phi_j(\bm{k}_j)$
describes the gap symmetry of the $j$ band in reciprocal space, which allows us to choose the gap form
depending upon each band as will be done in the following.
The vector potential ${\bm A}({\bm r})$ and the internal field ${\bm B}({\bm r})$ are related to 
${\bm B}({\bm r})=\nabla\times {\bm A}=\bar{\bm B}+{\bm b}({\bm r})$ with $\bar{\bm B}$ uniform field.

By following the same procedure by Eilenberger~\cite{eilenberger}, 
the functional derivatives with respect to $f_j=f(\omega_n,\bm{k}_j,\bm{r})$, and $f_j^{\dagger}=f^{\dagger}(\omega_n,\bm{k}_j,\bm{r})$
yield the so-called Eilenberger equation extended to the two band case:
\begin{equation}
\begin{split}
&\left\{\omega_n+i\mu B(\bm{r})+\bm{v}_j\cdot\left[\bm{\nabla }+i\bm{A}(\bm{r})\right]\right\}f_j=\Delta_j(\bm{k}_j, \bm{r})g_j,
\\
&\left\{\omega_n+i\mu B(\bm{r})-\bm{v}_j\cdot\left[\bm{\nabla }-i\bm{A}(\bm{r})\right]\right\}{f}^{\dagger}_j=\Delta_j^*(\bm{k}_j, \bm{r})g_j.
\end{split}\label{eq:Eilenberger}
\end{equation}
\noindent
This form is understandable because the fourth term in the free energy Eq.~\eqref{eq:freeenergy}, which includes the Green's functions,
is separable in the band index, thus the resultant equation of the functional derivative should be separable for each band.
The stationary conditions of Eq.~\eqref{eq:freeenergy} with respect to the functionals $\Delta_j^{\ast}(\bm{r})$ and the vector potential $\bm{A}(\bm{r})$ 
give rise to a complete set of the self-consistent equations extended to the two band case, which are given below; 
Eqs.~\eqref{eq:gap} and ~\eqref{eq:potential}. This complete set of the self-consistent equations 
coincides and is consistent with those obtained previously~\cite{koshelev,nakai2band}.


The electronic state is calculated by solving the Eilenberger equation Eq.~\eqref{eq:Eilenberger} in the clean limit~\cite{ichioka},
including the Pauli paramagnetic effect (PPE) due to the Zeeman term $\mu B(\bm{r})$~\cite{IchiokaPara}, 
where $\mu\!=\!\mu_{\rm B}B_0/\pi k_B T_{\rm c}$ 
is a renormalized Bohr magneton related to the so-called Maki parameter $\alpha_{\rm M}\!=\!1.76\mu$.
The quasiclassical Green's functions $g(\bm{k}_j,\bm{r},\omega_n\!+\!i\mu B)$, $ f(\bm{k}_j,\bm{r},\omega_n\!+\!i\mu B)$, and $f^{\dagger}(\bm{k}_j,\bm{r},\omega_n\!+\!i\mu B)$ with the band index $j$ depend on the direction of the Fermi momentum $\bm{k}_j$ for each band, the center-of-mass coordinate $\bm{r}$ for the Cooper pair, and Matsubara frequency $\omega_n\!=\!(2n\!+\!1)\pi k_{\rm B}T$ with $n\!\in\!\mathbb{Z}$.
They are calculated in a unit cell of the triangle vortex lattice.

%

The unit of Fermi velocity $v_{{\rm F}0}$ is defined by $N_{{\rm F}0}v_{{\rm F}0}^2\!\equiv\! N_{{\rm F}1}v_{{\rm F}1}^2\!+\!N_{{\rm F}2}v_{{\rm F}2}^2$, where the density of states (DOS) in the normal state at each Fermi surface is defined by $N_{{\rm F}0}\!\equiv\! N_{{\rm F}1}\!+\!N_{{\rm F}2}$.
Throughout this paper, temperatures, energies, lengths, and magnetic fields are, respectively, measured in units of the transition temperature $T_{\rm c}$, $\pi k_{\rm B} T_{\rm c}$, $\xi_0\!=\!\hbar v_{{\rm F}0}/2\pi k_{\rm B} T_{\rm c}$, and $B_0\!=\!\phi_0/2\pi\xi_0^2$. 
We calculate the spatial structure of $g$ in a fully self-consistent way.

The pairing potential $\Delta_j({\bf r})$ is calculated by the gap equation
\begin{align}
\Delta_j(\bm{r})=T\sum_{0<\omega_n\le\omega_{\rm c}}\sum_{l=1,2}V_{jl}N_{{\rm F}l}\left\langle (f_{l}+
{f}_{l}^{\dagger*})\phi_l(\bm{k}_l)\right\rangle_{\bm{k}_{l}},\label{eq:gap}
\end{align}
which is coupled via the interaction matrix $\hat{V}$.
We use the energy cutoff $\omega_{\rm c}\!=\!20k_{\rm B}T_{\rm c}$.
The vector potential is also self-consistently determined by
\begin{align}
\bm{\nabla }\!\times\!\bm{\nabla }\!\times\!\bm{A}\!=\!\bm{\nabla }\!\times\!\bm{M}_{\rm para}\!-\!\frac{T}{\kappa^2}\!\sum_{|\omega_n|\le\omega_{\rm c}}\!\sum_{j=1,2}N_{{\rm F}j}\!\left\langle\bm{v}_j{\rm Im}[g_j]\right\rangle_{\bm{k}_j}\!,\label{eq:potential}
\end{align}
which includes the contribution of the paramagnetic moment $\bm{M}_{\rm para}=(0,0,M_{\rm para})$ with
\begin{align}
M_{\rm para}\!=\!M_0\!\left(\!\frac{B(\bm{r})}{\bar{B}}\!-\!\frac{T}{\mu\bar{B}}\!\sum_{|\omega_n|<\omega_{\rm c}}\!\sum_{j=1,2}N_{{\rm F}j}\!\left\langle{\rm Im}[g_j]\right\rangle_{\bm{k}_j}\!\right)\!.\label{eq:Mpara}
\end{align}
Here $\bar{B}$ is the averaged flux density mentioned above, the normal state paramagnetic moment 
$M_0 = ({{\mu}}/{{\kappa}})^2 \bar{B} $, and 
${\kappa}=B_0/\pi k_{\rm B}T_{\rm c}\sqrt{8\pi N_{{\rm F}0}}$.   
The Ginzburg-Landau (GL) parameter $\kappa_{\rm GL}$ 
is the ratio of the penetration depth to coherence length for 
$\bar{\bf B}\parallel c$. 
Using Doria-Gubernatis-Rainer scaling~\cite{Doria}, we obtain the relation~\cite{ichioka} of $\bar{B}$ and the external field $H$.
%
%
The total magnetization $M_{\rm total}\!=\!\bar{B}\!-\!H$ including both the diamagnetic 
and the paramagnetic contributions is derived.

We solve Eq.~\eqref{eq:Eilenberger} with $i\omega_n\!\rightarrow\! E\!+\!i\eta$
for the electronic state.
The local density of states (LDOS) is given by $N_j(\bm{r},E)\!=\!N_{j,\uparrow }(\bm{r},E)\!+\!N_{j,\downarrow }(\bm{r},E)$ where
\begin{align}
N_{j,\sigma }(\bm{r},E)\!=\!N_{{\rm F}j}\!\left\langle{\rm Re}\left[g(\bm{k}_j,\bm{r},\omega_n\!+\!i\sigma\mu B)|_{i\omega_n\!\rightarrow\! E\!+\!i\eta }\right]\right\rangle_{\bm{k}_j}\!,
\end{align}
with $\sigma\!=\!1$ $(-1)$ for up (down) spin component.
We typically use the smearing factor $\eta\!=\!0.01$.
The DOS is obtained by the spatial average of the LDOS as $N(E)\!=\!\sum_jN_j(E)\!=\!\sum_j\langle N_{j,\uparrow }(\bm{r},E)\!+\!N_{j,\downarrow }(\bm{r},E)\rangle_{\bm{r}}$.

We consider a simplified model of a two-band system with a larger superconducting gap band  
(band-1) and a smaller gap band (band-2).
As a model of the Fermi surfaces, we use two quasi-two dimensional 
Fermi surfaces with rippled cylinder shapes. 
The Fermi velocity is assumed to be 
${\bf v}_j=(v_{j,a},v_{j,b},v_{j,c})\propto(\cos\phi,\sin\phi,\tilde{v}_{j,z} \sin k_{j,c})$ 
at 
${\bf k}_j=(k_{j,a},k_{j,b},k_{j,c})\propto(k_{j}\cos\phi, k_{j}\sin\phi,v_{j,c})$
on the Fermi surfaces~\cite{Hiragi}.  
We consider a case $\tilde{v}_{j,z}=1/\Gamma_j$,
to produce large anisotropy ratio of the coherence lengths, 

\begin{eqnarray}
\Gamma_j=\xi_{j,c} / \xi_{j,b} \sim  
\langle v_{j,c}^2 \rangle_{{\bf k}_j}^{1/2}/\langle v_{j,b}^2 \rangle_{{\bf k}_j}^{1/2}
\end{eqnarray} 

\noindent 
with $j=1,2$  where $\langle \cdots \rangle_{{\bf k}_j}$ 
indicates an average over the Fermi surface on each band. 
The magnetic field orientation is tilted by $\theta\equiv 90^{\circ}-\Omega$ from the $c$ axis 
towards the $ab$ plane. 
Since we set $z$ axis to the vortex line direction, 
the coordinate ${\bf r}=(x,y,z)$ for the vortex structure is related to the 
crystal coordinate $(a,b,c)$  
as $(x,y,z)=(a,b \cos\theta + c \sin\theta,c \cos\theta -b \sin\theta)$.  

We set unit vectors of the vortex lattice as

\begin{eqnarray} && 
{\bf u}_1=c({\alpha}/{2},-{\sqrt{3}}/{2}), {\bf u}_2=c({\alpha}/{2}, {\sqrt{3}}/{2})
\end{eqnarray} 
 
\noindent
with $c^2=2 \phi_0/ (\sqrt{3} \alpha \bar{B})$ and the vortex lattice anisotropy is defined by $\Gamma_{\rm VL}=\alpha/\sqrt3$.
The anisotropic ratio $\Gamma_j(\theta) \equiv
\xi_{j,y} / \xi_{j,x} \sim 
\langle v_{j,y}^2 \rangle_{{\bf k}_j}^{1/2}/\langle v_{j,x}^2 \rangle_{{\bf k}_j}^{1/2}$, which comes from the
Fermi velocity anisotropy,

\begin{eqnarray} && 
\Gamma_j(\theta) ={1\over{\sqrt{\cos^2\theta+\Gamma_j^{-2}\sin^2\theta}}}
\label{eq:anisotropy}
\end{eqnarray}

\noindent
with $j=1$ and 2.

To discuss $\bar{B}$-dependence of the internal field distribution 
${\rm B}({\bf r})=\nabla\times{\bf A}$,  
we consider flux line lattice (FLL) form factor 
${\bf F}({\bf q}_{h,k})=(F_{x(h,k)},F_{y(h,k)},F_{z(h,k)})$,  
which is obtained by Fourier transformation  
of the internal field distribution as  
${\bf B}({\bf r})=\sum_{h,k}{\bf F}({\bf q}_{h,k}) 
\exp({\rm i}{\bf q}_{h,k}\cdot{\bf r})$  
with the  wave vector ${\bf q}_{h,k}=h{\bf q}_1+k{\bf q}_2$.
$h$ and $k$ are integers. 
The unit vectors in reciprocal space 
are given by ${\bf q}_1=(2\pi/c)(1/\alpha,-1/\sqrt{3})$ 
and ${\bf q}_2=(2\pi/c)(1/\alpha,1/\sqrt{3})$.  
The $z$-component $|F_{z(h,k)}|^2$ from $B_z({\bf r})$ 
gives the intensity of conventional non-spinflip SANS. 
The transverse component, $|F_{\rm tr(h,k)}|^2=|F_{x(h,k)}|^2+|F_{y(h,k)}|^2$,  
is accessible by spin-flip SANS experiments.\cite{Kealey,Rastovski}

\section{Model construction and physics for $H\parallel c$}
\subsection{Specific heat and $H_{\rm c2}$ anisotropy ratio}

In order to determine the model parameters appropriate for Sr$_2$RuO$_4$,
we start out to fix the gap magnitudes $\Delta_{\gamma}(T)$ and $\Delta_{\beta}(T)$
and their nodal structures. We first analyze the electronic specific heat data $C(T)/T$ at
zero field~\cite{deguchi} by solving self-consistently the Eilenberger equation Eq.~\eqref{eq:Eilenberger} for the uniform system
without using the phenomenological so-called $\alpha$ model~\cite{alpha}.
Here we have assumed that the minor component induced  only by the Cooper pair transfer coupling $V_{12}$.
The direct attractive coupling among the minor component is vanishing; $V_{22}=0$.

\begin{figure}
\begin{center}
\includegraphics[width=5cm]{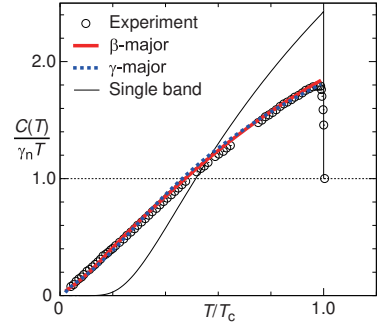}
\end{center}
\caption{\label{fig2}(Color online)
The specific heat data (open symbols)~\cite{deguchi} analysis by the $\beta$ major scenario (bold line) 
with $\Delta_{\beta}(0)/\Delta_{\gamma}(0)=2.5$
where the line node (full) gap is on the $\gamma$ ($\beta$) band 
and by  the $\gamma$ major scenario (dotted line) with $\Delta_{\gamma}(0)/\Delta_{\beta}(0)=1.7$
where the both bands contain the line nodes.
For comparison, the standard BCS case with the full gap is also shown (thin line). 
}
\end{figure}

 As seen from Fig.\ref{fig2} the $C(T)/T$ data is equally well explained either
 by 
 
 \noindent
 (A) the $\gamma$ scenario with $\Delta_{\gamma}(0)/\Delta_{\beta}(0)=1.7$
 at $T=0$ where both bands have line nodes as coincided with other authors\cite{nomura},

 \noindent
 (B) the $\beta$ scenario with $\Delta_{\beta}(0)/\Delta_{\gamma}(0)=2.5$
 where the  $\beta$ band has a full gap and the  $\gamma$ band line nodes.
 
 \noindent
 It is apparent that the linear $T$ behavior of $C(T)/T$ at lower $T$ indicates that the nodal gap is necessary somewhere;
 for the $\gamma$ scenario in both bands and for the the $\beta$ scenario only in the $\gamma$ band.
 This nodal structure difference between them is decisive  for the two scenarios as
 will be seen shortly.

\begin{figure}
\begin{center}
\includegraphics[width=5cm]{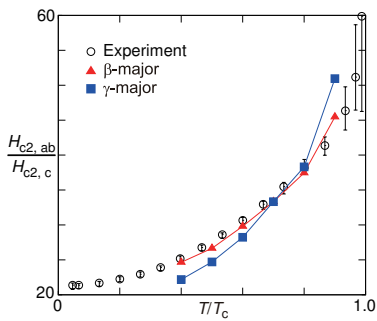}
\end{center}
\caption{\label{fig3}(Color online)
The upper critical field ratio $H_{{\rm c2},ab}(T)/H_{{\rm c2},c}(T)$ for the two directions
($H\parallel ab$ and $H\parallel c$). The two cases for the $\beta$ (filled triangles) and $\gamma$ scenarios (filled squares) are
compared with the experimental data (open symbols)~\cite{kittaka-ratio}.
This shows that the $\beta$ scenario is superior to the $\gamma$ scenario. 
}
\end{figure}

\begin{figure}
\begin{center}
\includegraphics[width=5cm]{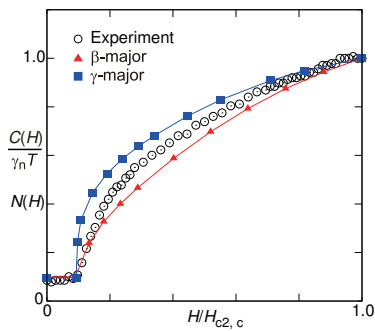}
\end{center}
\caption{\label{fig4}(Color online)
The comparison of the experimental data (open symbols)~\cite{deguchi} of the specific heat at $T$=60mK
as a function of $H(\parallel c)$ with the theoretical results (filled symbols) for the $\beta$ and $\gamma$ scenarios ($T=0.1T_{\rm c}$).
This shows too much low energy excitations released at lower fields in the $\gamma$ scenario
because the nodal gap exists in the minor band.
 }
\end{figure}

 According to Kittaka et al~\cite{kittaka-ratio}, the upper critical field ratio
 $H_{{\rm c2},ab}(T)/H_{{\rm c2},c}(T)$ for the two directions ($H\parallel ab$ and $H\parallel c$) is $T$-dependent,
 implying that PPE becomes stronger as the field applied to the $ab$ plane increases.
 Here we calculate $H_{{\rm c2},ab}(T)/H_{{\rm c2},c}(T)$ for both scenarios and depict the
 results in Fig. \ref{fig3}.
 Since near $T_{\rm c}$ the intrinsic effective mass anisotropy governs its tending limit ($T\rightarrow T_{\rm c}$), namely, 
 $H_{{\rm c2},ab}/H_{{\rm c2},c}\rightarrow$ 180 (60) for the  $\gamma$  ($\beta$) scenario.
 The experimental data~\cite{kittaka-ratio} shown support the $\beta$ scenario within the experimental accuracy
 where there is no indication of the ratio with tending to 180.
 This is one of the most clear signatures for the $\beta$ major scenario.
 By adjusting the $\mu$ parameters for both scenarios the best fittings are accomplished 
 in $\mu$=0.04 (0.02) for the $\beta$ ($\gamma$) scenario. 
 From now on we use those values in the following calculations.

 In order to further distinguish between the two scenarios, we
 take up the experimental data~\cite{deguchi} of the field dependence of $C(H)/T$
 for $H\parallel c$ at the available lowest temperature $T$=60 mK, which best mimics the
 zero-energy density of states $N(H)$ at $T=0$ in the theoretical calculation.
 In Fig. \ref{fig4} we compare the experimental data with the theoretical values
 as a function of the field applied parallel to the $c$ axis.
 As seen from it, the $\gamma$  scenario overestimates the data 
 while the $\beta$ scenario underestimates it. If taking into the finite $T$ effect
 in the experimental data, the theoretical curves should move up when considering
 thermal excitations,
 resulting in further departure of the  $\gamma$-major curve while
 the $\beta$-major curve comes closer to the data. 
 We also note from Fig. \ref{fig4} that too much low energy excitations are 
released at lower fields in the $\gamma$ scenario
because the nodal gap exists in the minor band, causing the overestimate.
This fact, which has been unnoticed so far is quite fatal for the $\gamma$ scenario.

 Thus it is clear from those two criteria
 that the $\beta$ scenario is far better than the $\gamma$ scenario.
 The best fitting also yields the $\kappa_{\rm GL}$ value for $H\parallel c$,
 that is, $\kappa_{\rm GL}$=2.7 whose value is used in the following computations.
 But we should keep in mind that this low $\kappa_{\rm GL}$ value delicately
 depends on the particular sample used. This quantity is known to be sample dependent.

 \subsection{Form factors for $H\parallel c$}

\begin{figure}
\begin{center}
\includegraphics[width=8.5cm]{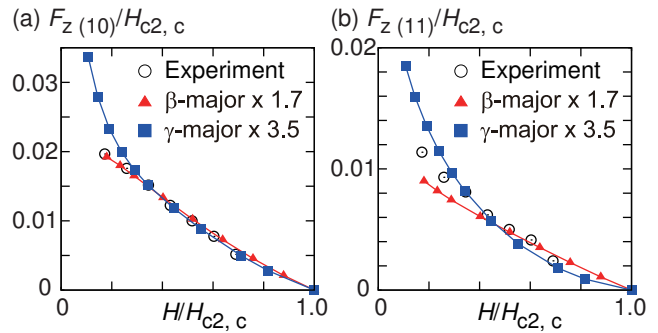}
\end{center}
\caption{\label{fig5}(Color online)
The longitudinal form factors $F_{z(10)}$ and $F_{z(11)}$ as a function of $H (\parallel c)$.
The experimental data (open symbols)~\cite{forgan} are compared with the two scenarios.
The theoretical results (filled symbols) are multiplied by a factor 1.7 (3.5) for the $\beta$ ($\gamma$)
case ($T=0.1T_{\rm c}$). This shows that the $\beta$ scenario is superior to the $\gamma$ scenario.
}
\end{figure}

 The form factors of the longitudinal components $F_{z(10)}$ and $F_{z(11)}$
 which are measured by SANS experiments~\cite{forgan} for $H\parallel c$ 
 are compared with the two scenarios whose magnitudes are multiplied by a factor of 1.7
 for the $\beta$ scenario and 3.5 for the $\gamma$ scenario.
 This is partly because the actual $\kappa_{\rm GL}$=2.7 determined by $C(H)/T$
 fitting previously might be different from the SANS experiment (In fact $H_{{\rm c2},c}$=58mT and 
 $\kappa_{\rm GL}$=2.0 differ from $H_{{\rm c2},c}$=75mT and 
 $\kappa_{\rm GL}$=2.3 in the best samples~\cite{MackenzieMaeno}).
Since the  form factor magnitudes are sensitive to the $\kappa_{\rm GL}$ value ($\propto \kappa_{\rm GL}^{-2}$),
it is permissible to adjust it to fit with data.
As seen from Fig. \ref{fig5} the field dependences of $F_{z(10)}$ and $F_{z(11)}$ are far better explained by the
$\beta$ scenario than by the $\gamma$ scenario.

\begin{figure}
\begin{center}
\includegraphics[width=8cm]{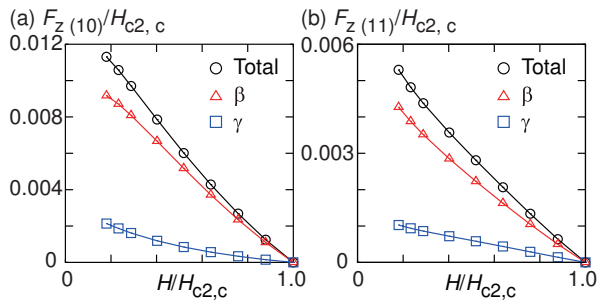}
\end{center}
\caption{\label{fig6}(Color online)
The longitudinal form factors $F_{z(10)}$ and $F_{z(11)}$ are decomposed into
the the  $\beta$ band contribution and $\gamma$ band contribution
in the case of the $\beta$ scenario ($T=0.1T_{\rm c}$). 
}
\end{figure}

In Fig. \ref{fig6} we decompose the longitudinal form factors $F_{z(10)}$ and $F_{z(11)}$
into the two contributions of the  $\beta$ band
and $\gamma$ band in the case of the $\beta$ scenario.
It is seen that the minor $\gamma$ contribution amounts to $\sim10\%$
of the total at this temperature $T=0.1T_{\rm c}$ for both $F_{z(10)}$ and $F_{z(11)}$.
Thus in order to understand the field dependence of the form factors,
the multiband effect is essential, which further becomes clear later.
Note in passing as mentioned in Introduction on MgB$_2$
Cubitt et al~\cite{mortenmgb2} discover the additional contribution of the minor $\pi$ band
 to the main $\sigma$ band contribution at lower $H$.
 This general trend here supports their discovery (see Fig. 2 in Ref. \onlinecite{mortenmgb2}).
 Note that the relative
 weight of the main $\beta$ contribution and minor $\gamma$ contribution depends 
 on temperature, field and field orientation $\Omega$. Generally as $T$ and $H$ decrease, the minor 
 $\gamma$ contribution increases because the two order parameters are more competitive there.

 \subsection{Phase diagrams in $H$ vs $\Omega$}

 Since we have determined the gap ratios for each scenario,
 it is possible to establish the phase diagrams on $H$ vs $\Omega$ plane.
 For $H\parallel c$ the two orbital limited $H^{i}_{{\rm c2},c}$ ratio ($i=\beta, \gamma$) 
 with the obvious notations is written as

\begin{eqnarray} && 
{H^{\beta}_{{\rm c2},c}\over H^{\gamma}_{{\rm c2},c}}={\xi^2_{\gamma,c}\over \xi^2_{\beta,c}}
=\left({\Delta_{\beta}\over \Delta_{\gamma}}\right)^2\cdot\left({v_{\gamma,c}\over v_{\beta,c}}\right)^2.
\end{eqnarray} 

\noindent
It is known\cite{MackenzieMaeno} that the Fermi velocity ratio $v_{\gamma,c}/v_{\beta,c}=0.5$. Thus
$H^{\beta}_{{\rm c2},c}/H^{\gamma}_{{\rm c2},c}=2.5^2\times0.5^2\sim1.6$
for the $\beta$ scenario while $H^{\beta}_{{\rm c2},c}/H^{\gamma}_{{\rm c2},c}
=1.7^{-2}\times0.5^2\sim0.1$ for the $\gamma$ scenario,
those corresponding to Fig. 1(a) and Fig. 1(c) at $\Omega=90^{\circ}$ respectively.
On the other hand, at $\Omega=0^{\circ}$

\begin{eqnarray} && 
{H^{\gamma}_{{\rm c2},ab}\over H^{\beta}_{{\rm c2},ab}}={\Gamma_{\gamma}H^{\gamma}_{{\rm c2},c}\over \Gamma_{\beta}H^{\beta}_{{\rm c2},c}}
=3\times{H^{\gamma}_{{\rm c2},c}\over H^{\beta}_{{\rm c2},c}}
\end{eqnarray} 
 
 \noindent
This yields $H^{\beta}_{{\rm c2},ab}/H^{\gamma}_{{\rm c2},ab}=3/1.6\sim1.9$
for the $\beta$ scenario while 
$H^{\beta}_{{\rm c2},ab}/H^{\gamma}_{{\rm c2},ab}=3/0.1\sim30$
for the $\gamma$ scenario, thus giving rise to the situations depicted in 
Fig. 1(a) and Fig. 1(c) at $\Omega=0^{\circ}$ respectively.
Note that the crossing point A of the two orbital limited $H_{\rm c2}$'s 
in Fig. 1(a) is located at around $\Omega_{\rm A}\sim1^{\circ}$.

\section{Form factors and vortex lattice anisotropy}
\subsection{Form factors}

The form factors FF are  a sensitive and useful probe
measured by SANS in order to detect the field distribution in the vortex state.
The SANS experiments~\cite{Rastovski,morten} on Sr$_2$RuO$_4$ are performed.
They find the transverse components of FF as a function of $\Omega$ near the $ab$ plane.
Here we obtain FF by evaluating the field distribution via a self-consistent solution of Eq.~\eqref{eq:Eilenberger}.

\begin{figure}
\begin{center}
\includegraphics[width=8cm]{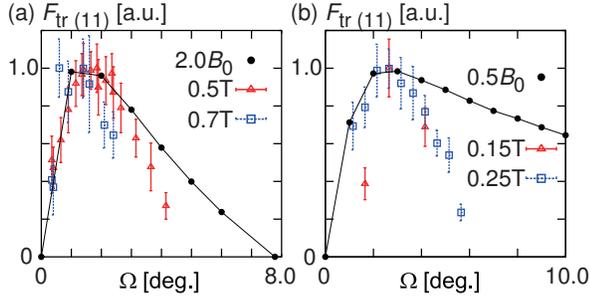}
\end{center}
\caption{\label{fig7}(Color online)
The transverse form factors $F_{\rm tr (11)}$ of the SANS data (open symbols)~\cite{Rastovski,morten} 
at $H$=0.7T and $H$=0.5T  are compared with
the theoretical results (filled symbols) for $B=2.0$ (a) and the data at  $H$=0.25T and 0.15T 
with $B=0.5$ ($T=0.5T_{\rm c}$) (b). 
}
\end{figure}

It is seen from Figs.~\ref{fig7} (a) and (b) that the angle $\Omega$ dependences of the transverse component of FF for
two fields $B=2.0$ (high field) and $B=0.5$ (low field) exhibit a different characteristic;
The high field data show a simple monotonic decrease after taking a maximum towards 
$\Omega=\Omega_{\rm c}$ at which the superconducting-normal state transition takes place.
The monotonic transverse FF curve is similar to that of the single band case shown in previous
papers\cite{amano1,amano2}. This high field scan ($B=2.0$) corresponds to the horizontal scanning path
in Fig.\ref{fig1}(b) where only the $\Gamma_{\beta}$  region is sensed.
The maximum position $\Omega_{\rm max}^{\rm FF}=0.9^{\circ}$ coincides with those 
of the single band result~\cite{amano2} with  $\Gamma=60$.
The overall features of the experimental data for $H=0.5$T and 0.7T are well reproduced as seen from Fig. \ref{fig7} (a)
except for a few data points at higher angles.

On the other hand, for the low field result ($B=0.5$) shown in Fig.\ref{fig7} (b), it is seen that the theoretical FF curve
coincides with the maximum angle of the experimental data at $H=0.15$T and 0.25T.
However, it deviates from those data after that.
The experimental data tend to vanish at round $\Omega_{\rm c}\sim 6^{\circ}$, which is by far from the 
known $\Omega_{\rm c}\sim30^{\circ}(18^{\circ})$ at $H=0.15$ (0.25)T, which should be ultimate vanishing angle for
FF amplitude. We will discuss this discrepancy shortly. 
Here we just point out that those low field data correspond to the low field scanning paths 
in Fig.\ref{fig1}(b) where the
crossover of the regions $\Gamma_{\beta}\rightarrow\Gamma_{\gamma}\rightarrow\Gamma_{\beta}$
is sensing with increasing $\Omega$.

\begin{figure}
\begin{center}
\includegraphics[width=8cm]{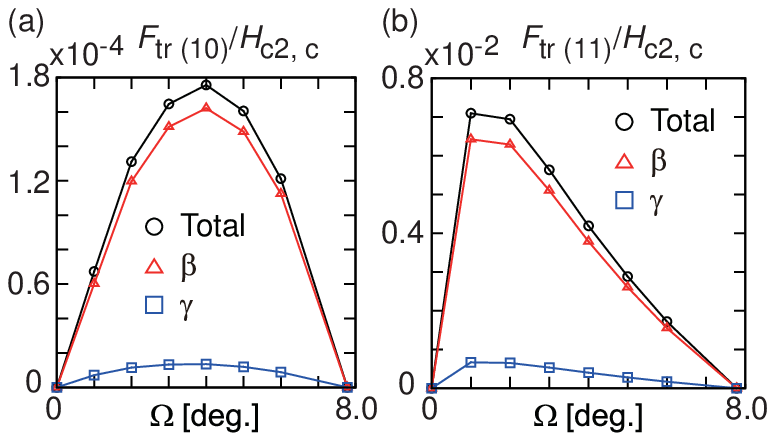}
\includegraphics[width=8cm]{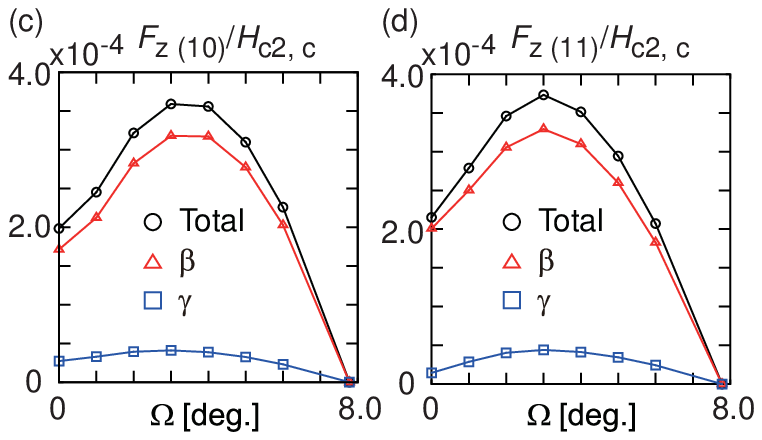}
\end{center}
\caption{\label{fig8}(Color online)
The decomposition of the transverse components of the form factors $F_{\rm tr (10)}$ (a)
and $F_{\rm tr (11)}$ (b) at high field $B=2.0$ ($T=0.5T_{\rm c}$).
The decomposition of the longitudinal components of the form factors $F_{\rm z (10)}$ (c)
and $F_{\rm z (11)}$ (d) at high field $B=2.0$ ($T=0.5T_{\rm c}$).
}
\end{figure}


\begin{figure}
\begin{center}
\includegraphics[width=8cm]{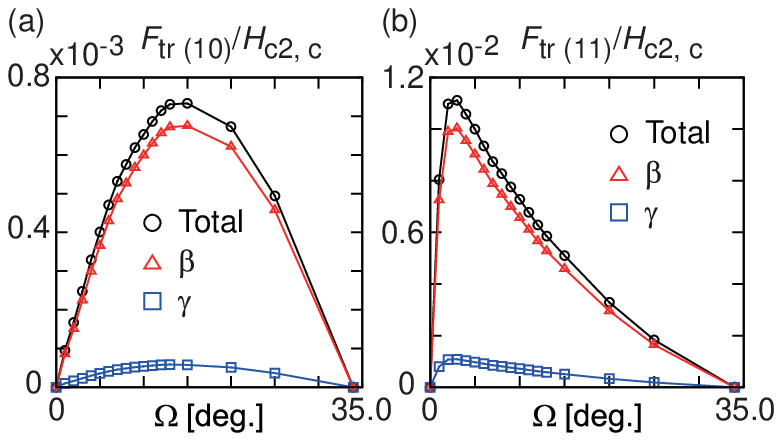}
\includegraphics[width=8cm]{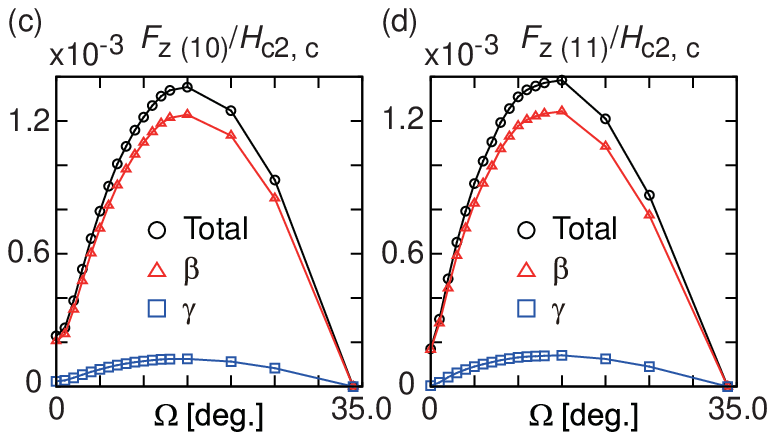}
\end{center}
\caption{\label{fig10}(Color online)
The decomposition of the transverse components of the form factors $F_{\rm tr (10)}$ (a)
and $F_{\rm tr (11)}$ (b) at low field $B=0.5$ ($T=0.5T_{\rm c}$).
The decomposition of the longitudinal components of the form factors $F_{\rm z (10)}$ (c)
and $F_{\rm z (11)}$ (d) at low field $B=0.5$ ($T=0.5T_{\rm c}$).
}
\end{figure}


We decompose the FF contributions from the major $\beta$ band and the minor $\gamma$ band
for the transverse components $F_{\rm tr(10)}$ (a) and $F_{\rm tr(11)}$ (b) and
the longitudinal components $F_{\rm z(10)}$ (c) and $F_{\rm z(10)}$ (d) as shown in Fig.~\ref{fig8} 
for the high field and Fig.~\ref{fig10} for the low field.
The relative weight of the minor component in this transverse FF is around $10\%$ of
the total FF and their peak contribution coincides with the maximum position $\Omega_{\rm max}$
for both $B=2.0$ (high field) and $B=0.5$ (low field) data.
As for the longitudinal components of $F_{\rm z(10)}$ and $F_{\rm z(10)}$, 
it is reasonable to see that the maximum angle  $\Omega_{\rm max}$
is shifted to a higher angle than $F_{\rm tr(11)}$.
This is because when $H_{\rm c2}(\Omega_{\rm c})$ is approached in which
the ``effective''  field virtually increases because the decreasing $\Delta(\Omega)$ means decreasing $H_{\rm p}$ 
and hence enhances PPE, thus pushing up the longitudinal FF $|F_z|$.

\begin{figure}
\begin{center}
\includegraphics[width=6cm]{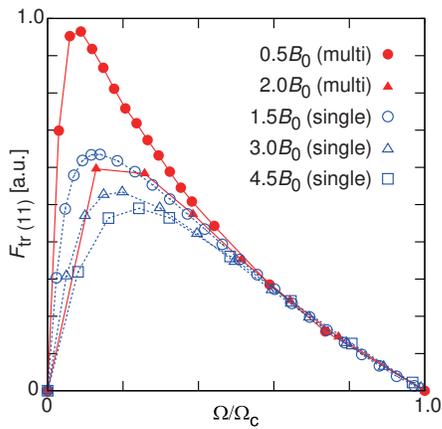}
\end{center}
\caption{\label{fig12}(Color online)
The comparison of the transverse form factors $F_{\rm tr (11)}$ with the single band results 
(open symbols)~\cite{amano2} for $B$=1.5, 3.0, and 4.5 
and the present multiband results (filled symbols) depicted in Fig. \ref{fig7} for $B$=2.0 and 0.5.
The vertical scale is adjusted so that 
the slopes of these curves near  $\Omega_{\rm c}$ coincide with each other.
It is clear that the low field theoretical result ($B=0.5$) behaves differently,
exhibiting a bimodal structure.
}
\end{figure}

In order to clearly see the different angle dependences $F_{\rm tr(11)}(\Omega)$ for the two cases
and to understand the discrepancy of the low field data mentioned above shown in Fig.~\ref{fig7} (b),
we replot those theoretical results normalized by the own $\Omega_{\rm c}$ in Fig.~\ref{fig12}
and compare those with the corresponding single band results~\cite{amano2}
where we have adjusted the vertical scale so that the slopes of these curves near  $\Omega_{\rm c}$ coincide with each other.
It is now seen clearly that
 
\noindent
(1) The high field result $B=2.0$ belongs to the single band universality curve.
This is because the high field scanning path is sensing only the $\Gamma_{\beta}$ region in Fig.~\ref{fig1} (b).
This is virtually same as in the single band case. Namely after taking the maximum the FF curve
simply goes to vanish at $\Omega_{\rm c}$.

\noindent
(2) The low field result $B=0.5$ behaves differently from those single band universality curves
and exhibits a ``bimodal'' $\Omega$ dependence where the FF peak is additionally enhanced.
Thus the curve just after taking the maximum tends to vanish earlier than at $\Omega_{\rm c}$
which is the ultimate vanishing angle.

Let us come back to understand the low field FF data shown in Fig.~\ref{fig7} (b).
The bimodal  $F_{\rm tr(11)}(\Omega)$ structure at the low field shown in Fig. ~\ref{fig10} (b)
where the slope just after the maximum differs
from the slope	at higher angles as $\Omega$ increases.
This crossover angle $\Omega_{\rm cross}\sim 4^{\circ}-5^{\circ}$. This low field scan
corresponds to the low field scanning path in Fig. 1(b) where the
crossover of the regions $\Gamma_{\beta}\rightarrow\Gamma_{\gamma}\rightarrow\Gamma_{\beta}$
is sensing with increasing $\Omega$.
In particular, the central peak $\Omega$ region ($1.0^{\circ}<\Omega<4.0^{\circ}$)
in the bimodal structure corresponds to the middle $\Gamma_{\gamma}$ region in this crossover.
This theoretical curve is compared with the SANS data~\cite{morten} at $H=0.25$T in Fig.\ref{fig7} (b). 
Although the present experimental  data do not exhibit this bimodal FF structure,
the existing data are understood as coming only from the middle $\Gamma_{\gamma}$
region among the bimodal structure because the extrapolated vanishing angle $\Omega_{\rm c}\sim7^{\circ}$.
This critical angle is too small compared with $\Omega_{\rm c}=18^{\circ}$ which must be the ultimate vanishing angle of the
FF intensity. Thus we interpret this so that the existing FF data are coming for the peak 
$\Gamma_{\gamma}$ region of the bimodal FF distribution.
This interpretation is strengthened later when analyzing the vortex lattice anisotropy $\Gamma_{\rm VL}$.

\subsection{Vortex lattice anisotropy $\Gamma_{\rm VL}$}

\begin{figure}
\begin{center}
\includegraphics[width=9cm]{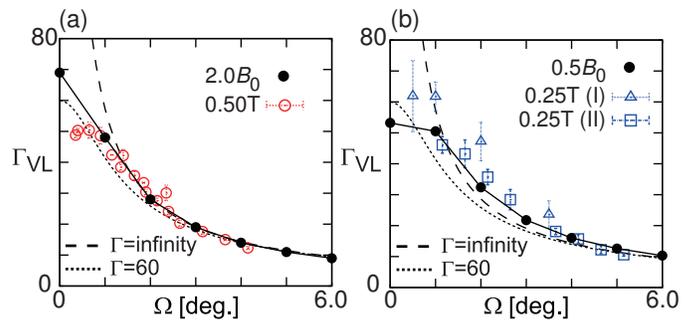}
\end{center}
\caption{\label{fig13}(Color online)
The comparison of the experimental data (open symbols)~\cite{Rastovski,morten} for $\Gamma_{\rm VL}(\Omega)$ for $H=0.5$T
with the high field results of $B=2.0$ (filled symbols) (a) and 
$H=0.25$T with the low field results of $B=0.5$ (filled symbols) (b) ($T=0.5T_{\rm c}$). 
The curves are drawn by the
effective mass formula Eq.~\eqref{eq:anisotropy} with $\Gamma=60$ (dotted line) and $\Gamma=\infty$ (dashed line).
}
\end{figure}

\begin{figure}
\begin{center}
\includegraphics[width=7cm]{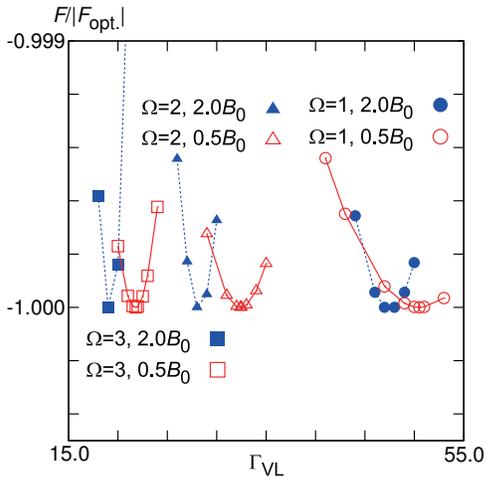}
\end{center}
\caption{\label{fig14}(Color online)
The free energy curves for various angles $\Omega$ as a function of $\Gamma_{\rm VL}$
for $B$=2.0 (filled symbols) and $B$=0.5 (open symbols) in order to emphasize that our computations are performed accurately 
enough to resolve the subtle differences of the $\Gamma_{\rm VL}$ values.
}
\end{figure}

Figure~\ref{fig13} shows the results of the vortex lattice anisotropy $\Gamma_{\rm VL}$
compared with the SANS experiments~\cite{morten}
where for a given $B$ and $\Omega$ the self-consistent solutions of Eq.~\eqref{eq:Eilenberger} 
are optimized for varying 
$\Gamma_{\rm VL}$ as shown in Fig.~\ref{fig14} to seek the free energy minimum
where the free energy form is given by Eq.~\eqref{eq:freeenergy}.
In general the optimized $\Gamma_{\rm VL}$ is that given 
neither by the effective mass anisotropies $\Gamma_{\beta}(\Omega)$ with $\Gamma_{\beta}$=60 
nor $\Gamma_{\gamma}(\Omega)$ with $\Gamma_{\gamma}$=180 alone
given by Eq.~\eqref{eq:anisotropy}.
This is obvious because those two anisotropies are coupled and competed by the
multiband effect.

The higher field results in Fig.~\ref{fig13} (a) follow rather well those given by the above formula of Eq.~\eqref{eq:anisotropy}
for $\Gamma_{\beta}(\Omega)$ for the higher angles $\Omega>3^{\circ}$.
We note that at $\Omega=0^{\circ}$ where $\Gamma_{\rm VL}(0^{\circ})$=70
corresponds precisely to the anisotropy $H_{\rm c2}^{\beta}(0^{\circ})/H_{\rm c2}^{\beta}(90^{\circ})$
for $T=0.5T_{\rm c}$.
Between the angles $1^{\circ}<\Omega<3^{\circ}$ the theoretical results deviate upwards, 
that is,  $\Gamma_{\rm VL}(\Omega)>\Gamma_{\beta}(\Omega)$. 
The high field scan in Fig.1(b) is barely touching the $\Gamma_{\gamma}$ region 
whose anisotropy can be certainly  larger than $\Gamma_{\beta}$=60 because $\Gamma_{\gamma}$=180.

The low field theoretical results shown in  Fig.~\ref{fig13} (b) follow the $\Gamma_{\beta}(\Omega)$ curve 
for  $0^{\circ}\leq\Omega<1^{\circ}$ and remarkably 
exceed the $\Gamma\rightarrow\infty$ curve in $1^{\circ}<\Omega<5^{\circ}$.
This window in $\Omega$ corresponding to the $\Gamma_{\gamma}$ region
appears because there the $\Gamma_{\gamma}$=180 anisotropy further modifies
the  $\Gamma_{\rm VL}$ value upwards, simultaneously enhancing the transverse FF.
We point out a theoretical fact that as $\Gamma_{\rm VL}(\Omega)$ increases, $F_{\rm tr}(\Omega)$
becomes larger. We anticipate that the results at lower temperatures than the present one at $T=0.5T_{\rm c}$
would improve the quantitative fittings of $\Gamma_{\rm VL}$ and simultaneously $F_{\rm tr}(\Omega)$.
Note that $\Gamma_{\rm VL}(\Omega=0^{\circ})$=52 corresponds to the single band anisotropy 
for the $\beta$ band at lower fields~\cite{amano2}.
The $H=0.25$T data nicely fit our theoretical result.
The important point here is that not only the experimental data exceed the $\Gamma\rightarrow\infty$
line, but also there exists a wider window $1^{\circ}<\Omega<5^{\circ}$ 
where the experimental data deviate from the single band $\Gamma_{\beta}(\Omega)$ curve.

We demonstrate in Fig.~\ref{fig14} that the optimal  $\Gamma_{\rm VL}$ is determined careful enough, which can
be larger than the corresponding $\Gamma\rightarrow\infty$ case in Eq.\eqref{eq:anisotropy}.
For example, for our $B$=0.5 case at $\Omega=2^{\circ}$ our $\Gamma_{\rm VL}$=32
while $\Gamma_{\rm VL}$=29 for $\Gamma={\infty}$ and $\Gamma_{\rm VL}$=28 for $\Gamma_{\beta}=60$.
Thus it can be said that the multiband effect helps enhancing $\Gamma_{\rm VL}$ beyond $\Gamma={\infty}$.

\subsection{Order parameters, free energy and magnetization}

\begin{figure}
\begin{center}
\includegraphics[width=5cm]{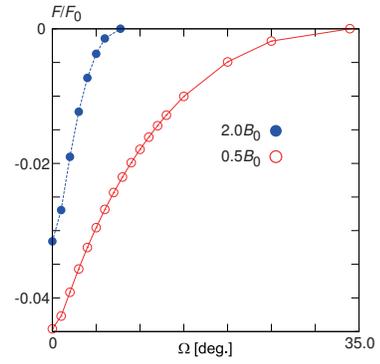}
\end{center}
\caption{\label{fig15}(Color online)
The angle dependences of the free energy for $B$=2.0 (filled symbols) and $B$=0.5 (open symbols),
showing a smooth change of the second order transition at $\Omega_{\rm c}$
where $\Omega_{\rm c}=7.78^{\circ}$ and $\Omega_{\rm c}=33.96^{\circ}$ respectively.
}
\end{figure}

We illustrate the $\Omega$ dependences of several physical quantities of interest
which are the basis of the form factors and magnetic torque calculations as shown shortly.
The free energy $F(\Omega)$ is shown in Fig.~\ref{fig15}, from which we will evaluate the magnetic torques.
Since the transition at $\Omega_{\rm c}$ is of second order at those fields and temperature (T=0.5T$_{\rm c}$),
$F(\Omega)$ becomes zero smoothly at the transition point $\Omega_{\rm c}$.

\begin{figure}
\begin{center}
\includegraphics[width=5cm]{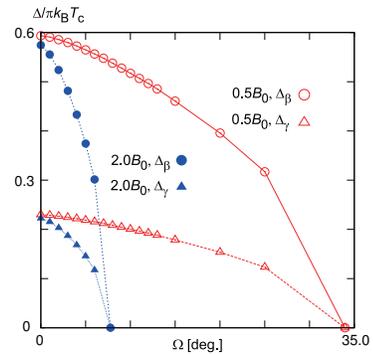}
\end{center}
\caption{\label{fig16}(Color online)
The angle dependences of the order parameter amplitudes $\Delta_{\beta}(\Omega)$ and
$\Delta_{\gamma}(\Omega)$ at the center of the vortex unit cell for $B$=2.0 (filled symbols) and $B$=0.5 (open symbols).
}
\end{figure}

As seen from Fig.~\ref{fig16}, the two order parameters; the major $\Delta_{\beta}(\Omega)$
and minor $\Delta_{\gamma}(\Omega)$ start decreasing from $\Omega=0^{\circ}$ towards 
$\Omega_{\rm c}$ as a function of $\Omega$. The two curves change smoothly 
in parallel because the minor component $\Delta_{\gamma}(\Omega)$ is induced by the major one
$\Delta_{\beta}(\Omega)$ through the Cooper pair transfer $V_{12}$ 
in the absence of  $V_{22}$. In this case we expect no hidden first order transition phenomena\cite{tsutsumi}
where there is an abrupt change of the two order parameters inside the superconducting state.

\begin{figure}
\begin{center}
\includegraphics[width=8cm]{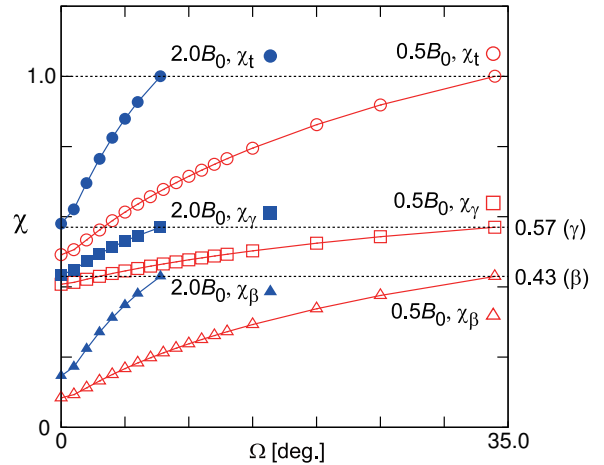}
\end{center}
\caption{\label{fig17}(Color online)
The angle dependences of the paramagnetic susceptibility $\chi_{\rm t}(\Omega)$
for $B$=2.0 (filled symbols) and $B$=0.5(open symbols), which are decomposed into $\chi_{\beta}(\Omega)$
and $\chi_{\gamma}(\Omega)$.
}
\end{figure}

In Fig.~\ref{fig17} the total paramagnetic susceptibility $\chi_{\rm t}(\Omega)$ and decomposed 
$\chi_{\beta}(\Omega)$ and $\chi_{\gamma}(\Omega)$ are displayed as a function of $\Omega$.
Since we assume the density of states $N_{{\rm F}\gamma}=43\%$ and $N_{{\rm F}\beta}=57\%$ of the total DOS,
the corresponding paramagnetic values $\chi_{\gamma}(\Omega)>\chi_{\beta}(\Omega)$ as expected.
We also point out that the base paramagnetic moment values at $\Omega=0^{\circ}$
 is large because the calculations are done at rather 
 high temperature $T=0.5T_{\rm c}$.
Needless to say, at T$\rightarrow0$ $\chi_{\rm t}$ should vanish at $\Omega=0^{\circ}$.
As $\Omega$ increases $\chi_{\rm t}(\Omega)$ becomes larger because the system is 
approaching the transition point at $\Omega_{\rm c}$ where the normal value $\chi_{\rm N}$
must be recovered, that is, $\chi_{\rm t}(\Omega_{\rm c})=\chi_{\rm N}$.

\section{Magnetic torques}

Magnetic torque $\tau(\Omega)$ defined by $\tau(\Omega)=\partial F(\Omega)/\partial \Omega$
provides several important information on a uniaxial anisotropic superconductor. 
We can know the intrinsic anisotropy of a system by the peak position of the torque curve 
$\tau(\Omega)$.
This can be easily performed by using a phenomenological theory based on London theory\cite{kogan}
for single band superconductors. In our previous papers\cite{amano1,amano2} we examine
the applicability of this approach and propose a modification to this.
Extending this single band Eilenberger calculations, here we show the results of the torque curves $\tau(\Omega)$
 for the present two band model and use those to analyze the data on Sr$_2$RuO$_4$ 
 where the torque curves are measured recently\cite{kittaka}.
 
\begin{figure}
\begin{center}
\includegraphics[width=8cm]{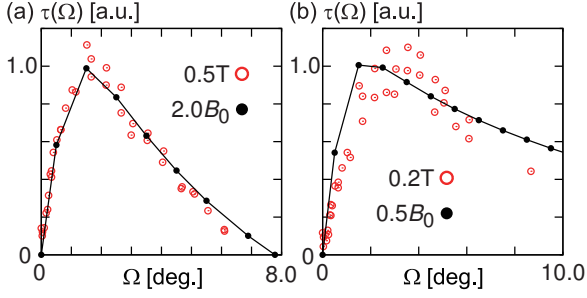}
\end{center}
\caption{\label{fig18}(Color online)
The angle dependences of the torques (filled symbols) for $B$=2.0 together with the experimental data (open symbols)~\cite{kittaka} 
of $H$=0.5T (a) and $B$=0.5 with the data of $H$=0.2T (b)
}
\end{figure}

 We first display our results for the torque curves in Fig.~\ref{fig18} where our results for $B=2.0$
 and $B=0.5$ are compared with the experimental data for $H$=0.5T (former) and $H$=0.2T (latter).
 It is seen from Fig.~\ref{fig18} (a) at the high field $B=2.0$ that the fitting is done well, 
 such as the peak position and the $\Omega_{\rm c}$ value.
 Since the high field result is sensing only on the $\Gamma_{\beta}$ region, it is reasonable that 
 the theoretical curve nicely explains the experimental data at $H$=0.5T and
 the peak position $\Omega^{\rm torque}_{\rm peak}=1.5^{\circ}$ coincides with the single band result~\cite{amano2}. 
 This peak position $\Omega^{\rm torque}_{\rm peak}$ also coincides with  $\Omega^{\rm FF}_{\rm peak}$
 of the form factor ($B=2.0$).
 Thus the four values $\Omega^{\rm torque}_{\rm peak}$ and 
 $\Omega^{\rm FF}_{\rm peak}$ both for theory and experiment are coinciding with each other, 
 leading us to firmly conclude that in high fields the system is virtually in the single band-like 
 $\Gamma_{\beta}$ region.

\begin{figure}
\begin{center}
\includegraphics[width=6cm]{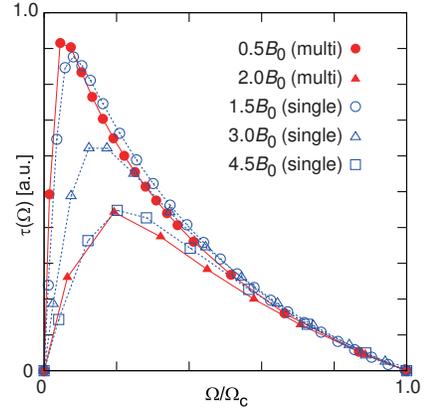}
\end{center}
\caption{\label{fig19}(Color online)
The torque curves as a function of $\Omega$ for $B$=2.0 and $B$=0.5 at $T=0.5T_{\rm c}$.
Those are compared with the single band results~\cite{amano2} for $B$=1.5, 3.0, and 4.5.
The vertical scale is adjusted so that 
the slopes of these curves near  $\Omega_{\rm c}$ coincide with each other.
}
\end{figure}

 On the other hand, the low field result ($B=0.5$) shown in Fig.~\ref{fig18} (b) explains the experimental 
 torque curves for $H$=0.2T, but the peak position differs slightly from each others.
 We also notice here that the theoretical $\Omega^{\rm torque}_{\rm peak}$$\neq$$\Omega^{\rm FF}_{\rm peak}$ 
 and $\Omega^{\rm FF}_{\rm peak}$ ($\sim 4^{\circ}$) agrees with the experimental data as
 shown before.

 It is interesting to compare our torque curves with those for the single band case which
 are displayed in Fig.~\ref{fig19}. It is seen that the two torque curves nicely correspond to the single band curves, which	
 is in contrast with the FF case shown in Fig.~\ref{fig12} where the low field curve markedly deviates
 from the single band case. This means that the magnetic torque is a 
 rather insensitive probe to see the subtle, but important multiband effect.
 In other words, the form factor measurement is sensitive enough to distinguish the multiband effect
 from the single band effect. This is because the torques $\tau(\Omega)=\partial F(\Omega)/\partial \Omega$
 comes from the total free energy $F(\Omega)$ while the FF is probing the particular Fourier component of the spatially
 modulated magnetic field in the mixed state selectively. 
 Thus it is natural to expect that the FF measurement is more sensitive than the torque measurement
 in picking up the multiband effect.

\begin{figure}
\begin{center}
\includegraphics[width=9cm]{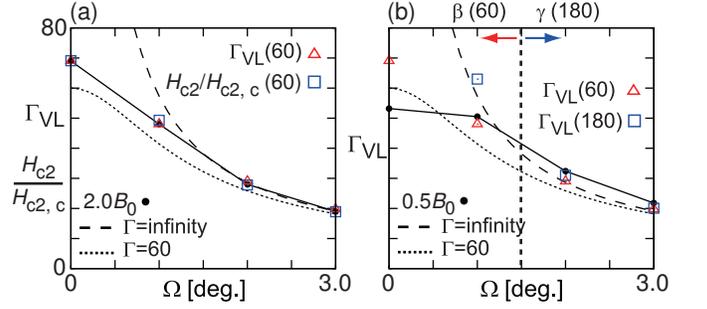}
\end{center}
\caption{\label{fig20}(Color online)
The comparison of the $\Gamma_{\rm VL}$ with the single band cases $\Gamma_{\rm VL}(60)$
and $\Gamma_{\rm VL}(180)$ and also with the $H_{\rm c2}$ anisotropy $H_{\rm c2}(\Omega)/H_{\rm c2,c}$.
(a) High field case $2.0$  where those three data $\Gamma_{\rm VL}$, $\Gamma_{\rm VL}(60)$ and 
$H_{\rm c2}/H_{\rm c2,c}$ coincide with each other, and (b) low field case $0.5$
where the lower (high) angle region corresponds to the $\beta$ ($\gamma$) band anisotropy.
The vertical dashed line at around $\Omega=1.5^{\circ}$ denotes its boundary.
The dotted (dashed) curves indicates the angle dependence of the effective mass formula Eq.~\eqref{eq:anisotropy}
for $\Gamma=60$ $(\Gamma=\infty)$.
}
\end{figure}

\section{Discussions}

\subsection{single band vs multiband}

In order to describe the transverse components of FF and $\Gamma_{\rm VL}$ data~\cite{morten}
near the $ab$ plane which are the main themes of the paper, we have done the calculations based on both scenarios,
first focusing on the $H\parallel c$ physics to fix the model parameters. 
Here we critically examine those scenarios comparatively.
It is obvious that the multiband scenario superior than the single band scenario~\cite{amano2} because
the former includes the latter as a limiting case.
A question is how effectively the single band  scenario can describe those data
or how the multiband description is inevitable for the Sr$_2$RuO$_4$ physics.
As already demonstrated and also shown in Fig.~\ref{fig20}, the overall features of those data can be reproduced by the single band model 
with $\Gamma_{\beta}$=60 in the higher fields above $H>0.5$T in the $H$ vs $\Omega$ plane where there
is a little trace for the existence of the $\gamma$ band with $\Gamma_{\gamma}$=180 (see Fig.~\ref{fig20}(a)).
In this sense the single band picture is enough for this region.
In contrast, however, it is necessary to retain the $\gamma$ band contribution 
in addition to the major $\beta$ band contribution in the low field region below 0.5T (see Fig.~\ref{fig20}(b)).
The former contribution is hidden and not explicit where the crossover occurs around $\Omega=1.5^{\circ}$
as indicated in Fig.~\ref{fig20}(b).  It is necessary and indispensable
to take it into account for explaining the FF  and $\Gamma_{\rm VL}$ data~\cite{morten}.
This is a part of the reasons why our single band theory~\cite{amano2} is successful to understand the mixed
state properties of the present strongly uniaxial anisotropic superconductor.
Moreover, depending upon the physical quantities of interest, the multiband effect is not so
apparent even in the low field region. Namely, the torque curves are quite insensitive to the
presence of the minor band and all the theoretical torque curves collapse into a universal curve
by appropriate scaling as discussed before (see Fig.~\ref{fig19}).

\subsection{$\beta$ main scenario vs $\gamma$ main scenario }

The question on which band the gap is large is much discussed in 
connection with the pairing mechanism to stabilize the chiral p-wave 
state by many authors~\cite{zhito,nomura,kivelson,simon}. Here we take a different view to consider this question
through the analysis of the FF, $\Gamma_{\rm VL}$ and torques.
If the $\gamma$ band is major, this survives in the high field region of the $H$ vs $\Omega$ plane
over the minor  $\beta$ band.
Then the SANS experiment should detect it, manifesting itself in $\Gamma_{\rm VL}$,
that is, $\Gamma_{\rm VL}\rightarrow 180$
as $\Omega\rightarrow 0$. However, both high fields and low filed data shown in Fig.~\ref{fig13} indicate 
$\Gamma_{\rm VL}\sim60$ or so,
which is direct evidence that the $\beta$ band with $\Gamma_{\beta}=60$ is major.
This conclusion is also supported by the $H_{\rm c2}$ ratio $H_{{\rm c2},ab}(T)/H_{{\rm c2},c}(T)\rightarrow60$
when $T\rightarrow T_{\rm c}$~\cite{{kittaka-ratio}} because near $T_{\rm c}$ this ratio directly
reflects the intrinsic Fermi velocity anisotropy~\cite{miranovic}, or the coherent length anisotropy.
Thus those experimental data obviously reveal that the $\beta$ band is major 
with certainty.

\subsection{Cooper pair tunneling $V_{12}$ vs direct attraction $V_{22}$: Hidden first order}

In general the two band contains the three pairing parameters $V_{11}$ ($V_{22}$)
is the attractive interaction for the major (minor) band, and  $V_{12}=V_{21}$ is the Cooper pair tunneling term 
or proximity induced term.
Here we set $V_{22}$=0 in this paper because there is no
or little indication for $V_{22}\neq 0$, which causes the so-called hidden first order phenomena~\cite{tsutsumi}.
Namely at $H^{\ast}$ inside $H_{\rm c2}$ certain physical quantities exhibit  a sudden change as a
function of $H$, such as the Sommerfeld coefficient $\gamma (H)$ or the magnetization curve
as observed in CeCu$_2$Si$_2$~\cite{kittakace}, UBe$_{13}$~\cite{shimizu} and KFe$_2$As$_2$~\cite{kittaka122}.
This means that  $V_{22}\sim 0$ in Sr$_2$RuO$_4$ and the minor  $\gamma$ band pairing is 
exclusively induced by the major $\gamma$ band through $V_{12}$.

\subsection{Expected low $T$ behaviors}

Because of several technical reasons our main computations have been done at relatively high $T$,
namely $T=0.5T_{\rm c}$, which are nevertheless successful in capturing the characteristic 
features in the FF, $\Gamma_{\rm VL}$ and torques in a qualitative level.
To understand those data in a quantitative level it is necessary to go into lower $T$ which is demanding 
computationally. Here we anticipate possible outcomes if performing it.
As shown in Fig.~\ref{fig7} we have mentioned that the low filed FF ($B$=0.5) differs 
from that in $B$=2.0, which belongs to the single band universality class.
The FF in low angle $\Omega$ is enormously enhanced by the assistance
from the minor band which amounts almost 10$\%$ even at $T=0.5T_{\rm c}$.
In low $T$ calculation this enhancement of the FF should increase
and the FF angle dependence becomes more similar to the data shown in Fig.~\ref{fig7} (b).
This expectation is reasonable because at lower $T$ the induced paramagnetic moments
are confined in the vortex core, making more contrast in the spatial field distribution
and thus enhancing the FF amplitude. See also Fig.~\ref{fig17} where the paramagnetic susceptibility or 
paramagnetic moments of the minor $\gamma$ band are
almost exhausted at $\Omega$=0 to its normal value, meaning that is not
confined in the core. Thus, the paramagnetic moments are spreading out the whole space uniformly.

\subsection{Other multiband superconductors with PPE}

We can deepen our understandings of the present material by comparing  
similar multiband superconductors, CeCu$_2$Si$_2$~\cite{kittakace}, UBe$_{13}$~\cite{shimizu} and KFe$_2$As$_2$~\cite{kittaka122}.
The first two  are heavy fermion materials known to have multiband with all full gaps
 belonging to the spin singlet
 category where $H_{\rm c2}$ is strongly suppressed and
 the hidden first order like anomalies at $H^{\ast}$ exist.
 Thus as mentioned $V_{22}$ is indispensable for those systems.
 According to the recent SANS experiment~\cite{kuhn} the typical multiband Fe pnictide 
KFe$_2$As$_2$ is similar to Sr$_2$RuO$_4$ in the point that the vortex
anisotropy $\Gamma_{\rm VL}$ observed differs from the $H_{\rm c2}$ anisotropy.
This can be also explained by PPE. Thus the present theoretical framework, which
is quite general  might be able to explain those systems.

\subsection{Gap structure: vertical line node vs horizontal line node}

We have assigned the nodal structure that the dominant $\beta$ band
is a full gap while the minor $\gamma$ is vertical line nodes.
In order to reproduce the square vortex lattice oriented to the 
(110) direction observed  for $H\parallel c$ which dominates the whole space 
measured in the $H$ vs $T$ phase~\cite{forgan}.
The gap (or near) nodes should be in this direction in reciprocal space~\cite{nakai}, implying a 
$d_{x^2-y^2}$ like nodal structure, contrary to the claim by Deguchi et al~\cite{deguchi} 
who measure the specific heat by rotating the
field direction and see the four-fold oscillation patterns whose minima are located
for the (100) direction. They conclude that the $\gamma$ band has a 
$d_{xy}$ like nodal structure.
However, this assignment is difficult to explain the square vortex lattice
orientated along the (110) direction for $H\parallel c$.
Even if we take into account the in-plane Fermi velocity anisotropy in the $\beta$ and $\alpha$ bands 
whose minima are oriented along the (100) direction~\cite{singh}, thus preferring the square lattice oriented along
the (100) direction as discussed previously~\cite{agterberg}.
Thus the best way to avoid this difficulty is to simply consider that the $\gamma$ band nodal 
structure is $d_{x^2-y^2}$ like.
The specific heat oscillation experiment done above 120mK is still too high
to see the sign changing of the oscillation pattern because the $\gamma$ band
is minor  where the expected sign changing temperature theoretically~\cite{Hiragi} 
and experimentally~\cite{an} usually located at 0.1$T_{\rm c}\sim 150$mK
must be lowered.

This kind of the band-dependent nodal structure differs from the idea based on the symmetry protected
nodal structure where the all plural bands are governed by the same gap symmetry.
From this point of view, it is possible that the $d_{x^2-y^2}$ nodal structure may not be truly 
sign-changing symmetry, rather an extended s-wave type with the anisotropic gap whose minima are along the (110) direction.

Some authors~\cite{hasegawa} assert that the horizontal line nodes compatible to a chiral triplet state $(p_x+ip_y)\cos p_z$.
From the present point of view, the Sommerfeld coefficient $\gamma(H)$ behavior for $H\parallel c$ may not be consistent
with this which gives a too larger $\gamma(H)$ at low field region than the experimental data~\cite{deguchi} shown in Fig.~\ref{fig4}.
In this connection we should mention that no one succeeded in explaining the
$\gamma(H)$ behavior~\cite{deguchi} for $H\parallel ab$. In particular, it is viewed that the initial rise of $H\parallel ab$ is often assigned 
to the $\alpha+\beta$ DOS because the plateau of $\gamma(H)$ for this mid field region 
$H\sim 0.4-0.5$T seems to correspond to $\alpha+\beta$ DOS (43$\%$ of the total), 
which was taken as supporting evidence of the $\gamma$ major scenario.
Since we cannot accept this view anymore, a full  understudying of the $\gamma(H)$ behavior for 
$H\parallel ab$ remains mystery.

\subsection{Perspectives----Future experiments; Knight shift and FFLO }

There remain certainly important experiments to put forth the research front of this interesting material:
NMR experiments have been extensively done~\cite{MackenzieMaeno}, showing the absence of any 
change of the Knight shift~\cite{knight} below $T_{\rm c}$.
 This is interpreted as freely rotatable d-vector in a spin triplet pairing, keeping always it perpendicular to
 an external field which is as small as $\sim$mT, a big mystery because according to the recent ARPES~\cite{ARPES} 
 the spin-orbit coupling which locks the d-vector to the crystalline lattice is an order of 200mV.
 Since the recent magnetization experiment~\cite{kittaka}  is directly able to detect the spin susceptibility change at the
 first order transition $H_{{\rm c2},ab}$  which amounts to $\sim10\%$ drop compared with the normal value.
 This is in sharp contrast with the Knight shift experiment by NMR which calls for reexamination of NMR experiments. 
 
 Here we propose the $T_1$ measurement to detect the FFLO state~\cite{fulde,larkin} expected 
 to exist along $H_{{\rm c2},ab}$ below $T\simeq$0.8K where the first order transition is observed.
Recently anomalous $T_1^{-1}$ enhancement is observed~\cite{vesna}  when entering the FFLO state due to the 
appearance of the zero energy state at the domain walls  where the FFLO order parameter is $\pi$-shifted. There is a 
good chance to observe it if the second phase below $H_{{\rm c2},ab}$ really exists, which we believe so.
A necessary condition is that $T_2$ is short enough
so that the spin-lattice relaxation $T_1$ process is dominated
through those zero energy states as in $\kappa$-(BEDT-TTF)$_2$Cu(NCS)$_2$ case~\cite{vesna}.
In this connection we mention a recent $\mu$SR experiment~\cite{muSR} that probes the peculiar vortex morphology 
at low fields of $H\parallel c$ and related theory based on hidden criticality associated with multi-bandness~\cite{milo}.

\section{Summary and conclusion}
Based on the microscopic Eilenberger theory extended to a multiband case, we have studied the mixed state properties
of a uniaxial anisotropic type II superconductor, focusing on the interplay between the Pauli paramagnetic
effect and multiband effect. A two band model calculation is set up and applied to Sr$_2$RuO$_4$.
We have succeeded in reproducing the data both
of the form factors of SANS experiments~\cite{Rastovski,morten} and magnetic torque experiment~\cite{kittaka}.
That leads us to the conclusion that to understand the physics in Sr$_2$RuO$_4$ 
it is indispensable to consider both the Pauli paramagnetic
effect and multiband effect simultaneously, which conspire to give rise to a variety of mysteries 
associated with the pairing symmetry determination in this material.

As agreed with the previous identification based on the single band analysis~\cite{amano2},
the pairing symmetry in Sr$_2$RuO$_4$ is either singlet which is most likely or triplet
with the d-vector locked in the $ab$ plane which is less likely.
The $\beta$ ($\gamma$) band is major (minor) with the mass anisotropy 60 (180).
Namely, the $\beta$ ($\gamma$) band  has a lager (smaller) gap. 
The gap structure is a full gap in the $\beta$ band
while in the $\gamma$ band it is $d_{x^2-y^2}$ like.
 This simple picture was difficult to reach because of the extreme two-dimensionality of
 this material which prevents conventional experimental access.
 Now the dedicated and refined experimental tools~\cite{Rastovski,kittaka,yonezawa1,yonezawa2} which are able to align the 
 magnetic field direction accurately within 1$^\circ$ enable us to uncover the
 physics of Sr$_2$RuO$_4$.

\begin{acknowledgments}
The authors thank M. R. Eskildsen, S. Kuhn, and C. Rastovski for providing us unpublished SANS data
and also M. Ichioka, M. Ishihara, and Y. Amano for discussing on the theoretical side, 
and S. Kittaka, A. Kasahara, T. Sakakibara, K. Ishida, Y. Maeno, S. Yonezawa and M. Takigawa
for discussing on the experiment side.
K. M. is supported by Grant-in-Aid for Scientific Research No. 26400360 and No. 25103716 
from the Japan Society for the Promotion of Science.
\end{acknowledgments}


\end{document}